\documentclass[epj]{webofc}
\usepackage[varg]{txfonts}   
\usepackage{hyperref}
\usepackage[utf8]{inputenc}
\wocname{EPJ Web of Conferences}
\woctitle{ICNFP 2017}
\begin{document}
\selectlanguage{english}
\title{Perspectives on the detection of  supersymmetric Dark Matter}
%
%

\author{Wim de Boer\inst{1}\fnsep\thanks{\email{wim.de.boer@kit.edu}} 
}

\institute{ Karlsruher Institute of Technology, Institute of Experimental Particle Physics, 76131 Karlsruhe, Germany 
}

\abstract{%
Up to now searches for Dark Matter (DM) detection have not been successful, either because our paradigm in how DM signals should look like are wrong or the detector sensitivity is still too low in spite of the large progress  made in recent years. We discuss both possibilities starting with what we know about DM from cosmology and why Supersymmetry provides such an interesting paradigm for cosmology and particle physics in order to appreciate what it means to give up this paradigm. In addition, we compare the   predicted cross sections for direct and indirect DM detection with observations with emphasis on the latest developments. Especially, we discuss the possible origins of the two hotly debated candidates for a DM annihilation signal, namely the positron excess and the Fermi GeV excess, which are unfortunately incompatible with each other and more mundane  astrophysical explanations exist.
}
\maketitle
\section{Introduction}
\label{intro}
With detailed satellite observations from our universe it has become clear that over 80\% of the matter is dark matter (DM), i.e. matter not visible by normal telescopes and behaving under gravity like visible matter, i.e. feeling attractive gravity. In contrast, the dark energy $\Lambda$ (DE) behaves different from matter, since it leads to repulsive gravity, as is evident from the accelerated expansion of the Universe, see Sect. 27 or Ref. \cite{Patrignani:2016xqp} for a review and references therein. We know all this from two basic observations (see standard textbooks on  cosmology, e.g. \cite{Kolb:1990vq,Roos:1994fz,Bergstrom:1999kd,Dodelson:2003ft,Ryden:2016gpw}: 1) The acoustic peaks in the cosmic microwave background (CMB), which is the remnant of the first light in the Universe. 
2) The accelerated expansion of the universe, i.e. the average distance between Galaxies increases in time, somewhat similar to a Helium balloon starting its journey into the sky. A satellite observer would contest that the balloon is subjected to repulsive gravity. He may be wondering how this can happen, if he does not know about the interactions between the balloon and its surroundings. 
Apart from these two basic observations the ''Standard Model of Cosmology'', called the $\Lambda$CDM model for the combination of $\Lambda$ and cold DM, has been confirmed by several independent observations, like baryonic acoustic oscillations, the power spectrum of the distribution of matter in the Universe, so its acceptance in the community starts to approach the acceptance of the Standard Model (SM) of Particle Physics. But the SM  is incomplete, since it cannot answer the questions concerning the nature of DM and DE, i.e. it describes only 5\% of the energy content of the universe. The properties of DM particles will be discussed  in Sect. \ref{dm}. For DM  the supersymmetric extension of the SM has an excellent  candidate, the so-called lightest supersymmetric particle (LSP), although  other candidates exist \cite{Bertone:2004pz}.

But Supersymmetry has many other advantages, as will be discussed  in Sect. \ref{susy}.  Therefore, we concentrate on supersymmetric DM.
The DM searches and interpretations are discussed in Sect. \ref{detection}. We will also discuss why the so-called Fermi excess in the gamma-ray sky and the AMS-02 positron excess, both discussed as signals for DM detection, can be easily (and better) explained by astrophysical origins, so they are no compelling (and especially not a consistent) argument for DM.  In the conclusion (Sect. \ref{conc})  we summarize why at present no DM signals have been observed, which is either because our paradigm of WIMP detection is wrong or because the experiments are not sensitive enough. In Supersymmetry the latter is possible, but future experiments will be able to cover a large region of the available parameter space. If still no DM signals are found with future experiments, the pressure to look for another paradigm will increase correspondingly.

 \begin{figure}[]
\centering
\includegraphics[width=7cm,clip]{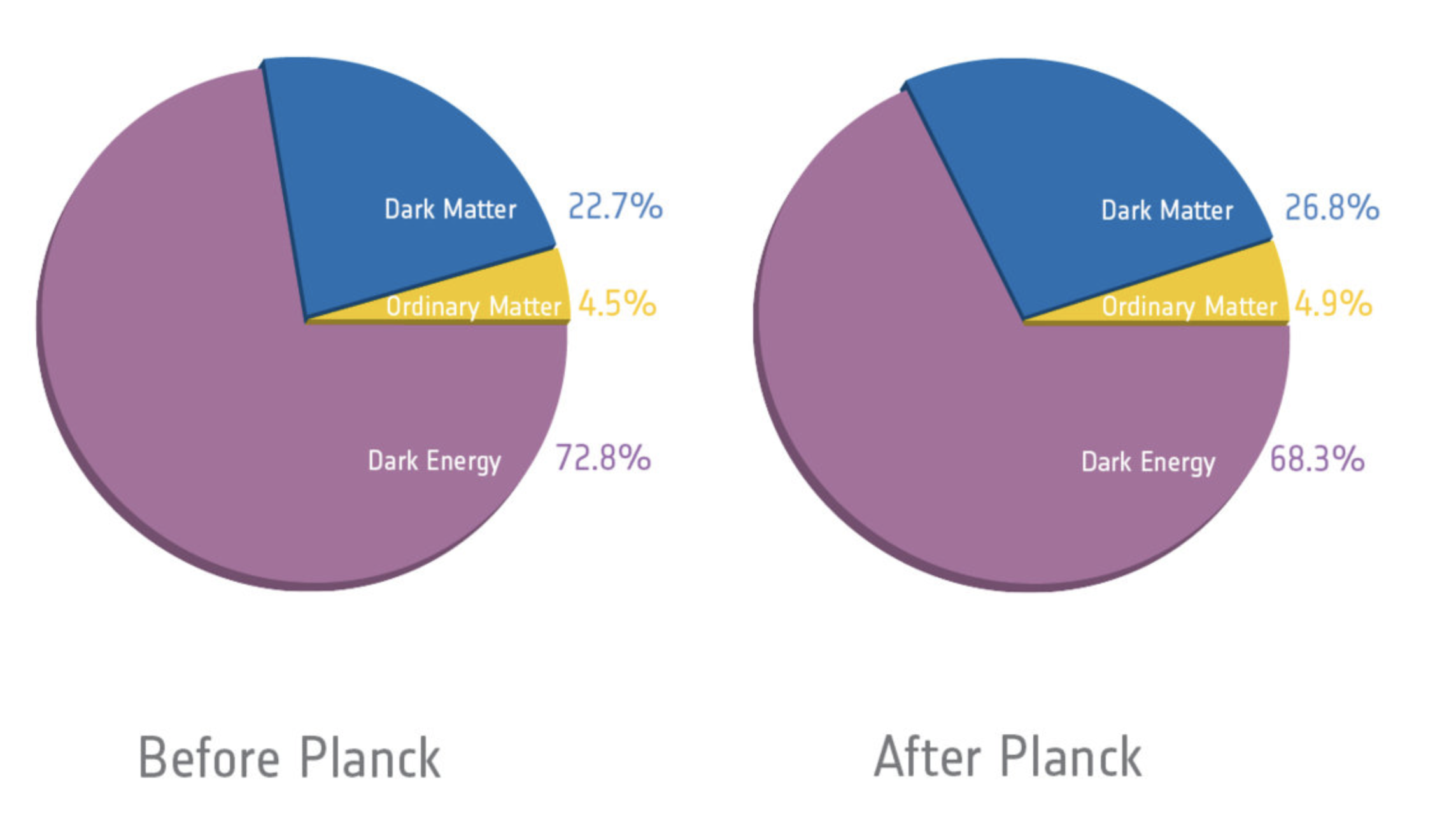}
\caption{The energy components of the universe. From the ESA webpage showing the results from the precise measurements by the Planck satellite, which increase the DM content 15\% in comparison with previous determinations. 
}
\label{f1}       
\end{figure}
\section{What do we know about DM?}
\label{dm}

The term ''dark matter'' was coined in 1932 by the Dutch astronomer Jan Oort, who studied the local matter density  from the movements of stars in the solar neighbourhood \cite{1932BAN.....6..249O}.  From these movements he understood (for the first time!) that the Sun may not  be the Galactic centre,but the Galaxy has a massive centre at a distance far away. This strongly contradicted the model of his thesis advisor, Prof. Kapteyn in Groningen, who considered the Sun to be the Galactic centre. He inferred that there may be dark matter in our Galaxy and calculated the ''Oort limit" of the local mass density, which he found to be 0.092 $M_\odot/pc^3$, while the mass in stars was estimated to be only  0.038 $M_\odot/pc^3$, so there must be dark matter. We now know, that most of this dark matter is coming from the gas  and up to 30\% from DM \cite{deBoer:2010eh}. In 1933 Zwicky studied the movements of galaxies in the Coma Cluster and discovered the need for  mass in addition to the visible mass  \cite{Zwicky:1933gu}. He called it DM too (independently).

Since  DM feels attractive gravity it presumably consists of particles, which have mass. Furthermore, DM particles have to be practically neutral, since else they would have been  detected by the ionization in charged particle detectors. Neutral particles can either have strong or weak interactions. If they would feel strong interactions, they would clump in the centre of the Galaxies, just like the baryons, which loose kinetic energy by energy losses, if they pass through the centre with its high baryon density. However, the rotation curves (rotation velocity of stars versus distance from the centre) do not fall off steeply outside the region of visible matter, which implies that there must be mass outside the radius of the visible matter. Since one does not see it, it must be DM. Hence, the DM consists of WIMPs (Weakly Interacting Massive Particles). In principle, WIMPs could be the neutrinos of the Standard Model (SM), which pass through the Galactic centre without loosing energy, thus moving in the gravitational field on Keplerian orbits reaching far outside the radius of visible matter. However, there are problems with neutrinos as DM candidate. 
The reason for the small neutrino contribution is simply the low neutrino mass, for which an upper limit of the sum of the three neutrino species  of 0.2 eV can be obtained from the power spectrum of the Galactic density fluctuations combined with other observations, see p.760 of Ref.  \cite{Patrignani:2016xqp}. The power spectrum shows that the small scale clustering of the Galaxies requires that DM must be non-relativistic in the early universe  (so-called cold DM), since for relativistic DM particles  the Jeans scale\footnote{This scale determines when the contraction by gravity is stronger than the expansion.} is too large. 
\begin{figure}[t]
\centering
\includegraphics[width=0.8\textwidth,clip]{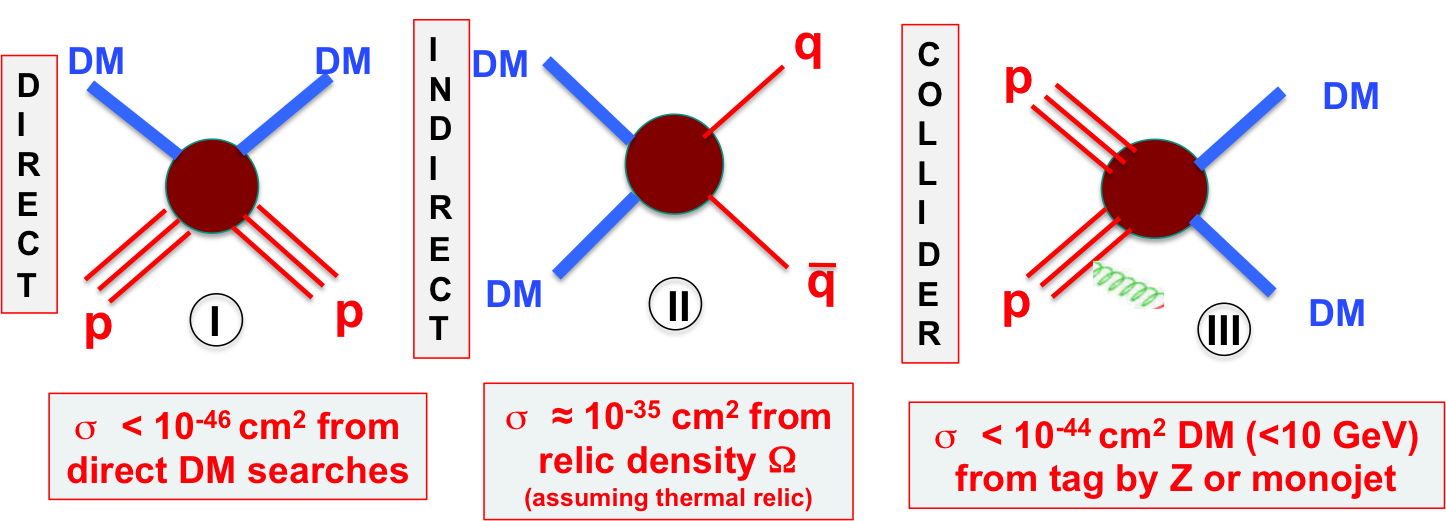}
\caption{The Feynman diagrams for WIMP interactions by direct searches , indirect searches and production at colliders.
}
\label{f2}       
\end{figure}

Assuming the WIMPs were created in the early, hot universe in thermal equilibrium with all other particles leads immediately to an estimate of the annihilation cross section.  One can ask: if particles are in thermal equilibrium at a high temperature, they fall out of equilibrium (''freeze-out'') at lower temperatures, when the lighter particles do not have enough kinetic energy to produce the more massive WIMPs. Then the  number density of WIMPs decreases exponentially by annihilation of WIMPs into lighter particles and one may wonder, why at present we still have a relic density.
In a static universe the DM density would indeed be zero in this scenario. However, in an expanding universe the expansion rate, given by the Hubble constant $H$, may become larger than the annihilation rate $\Gamma$, proportional to the annihilation cross section $\sigma_{anni}$. If $\Gamma\approx H$, the annihilation stops for a certain relic number density of WIMPs, which can be observed nowadays as the 23\% relic density, see   Fig. \ref{f1}.  So the present relic density and hence, the annihilation cross section is determined by the Hubble constant.  
Putting in numbers one finds for the velocity averaged annihilation cross section of the WIMP particles: $<\sigma_{anni} v>= 2.2\cdot10^{-26}$ cm$^3$/s for WIMP masses above 10 GeV and slightly higher for lower masses \cite{Steigman:2012nb}.  Comparing this annihilation cross section with the upper limit on the scattering cross section for a WIMP mass of 50 GeV, as observed in direct DM searches, leads to the result that these cross sections are more than 10 orders of magnitude apart, which means that if two WIMP particles hit each other, they have a 10 orders of magnitude higher probability to interact than if a WIMP particle hits a proton. This is surprising, since the basic Feynman diagrams are very similar, as shown in Fig. \ref{f2}. The blob indicates the exchange of energy via a neutral and weakly interaction particle, for which only the Z- or Higgs boson are candidates. The more than 10 orders of magnitude can be explained, if the Higgs boson is the dominant propagator, since the Higgs boson hardly couples to the light quarks inside a proton (or neutron)\footnote{Although the small energy transfer in p-WIMP scattering  suppresses the cross section by the t-channel propagator, it is not enough to explain the 10 orders of magnitude.}. The annihilation will then be predominantly into  heavy quarks or leptons, like b-quarks or tau-leptons, or if kinematics permits, top-quarks. Then the light quarks do not play any role. So if the WIMPs are thermal relics, we know already quite a bit: the annihilation- and cross sections are severely constrained by experimental data, which suggests that the interactions are dominated by Higgs exchange. This knowledge helps in drawing conclusions from future data. 

\begin{figure}[t]
\centering
\includegraphics[width=0.7\textwidth,clip]{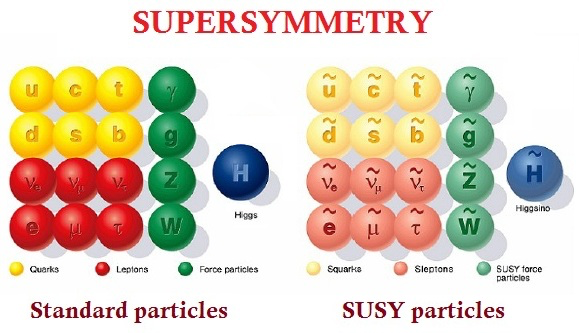}
\caption{The particle content of the SM (left) and MSSM (right). From CERN.
}
\label{f3}       
\end{figure}

\section{Why is Supersymmetry so interesting for DM, Particle Physics and Cosmology?}
\label{susy}
As mentioned before the SM has no good WIMP candidate, so one needs an extension of the SM. Supersymmetry is the most popular extension of the SM, since it solves several shortcomings of the SM simultaneously, especially it provides an excellent WIMP candidate. Details can be found in reviews \cite{Haber:1984rc,deBoer:1994dg,Martin:1997ns,Kazakov:2015ipa}.
In the SM the mediators of the strong, electromagnetic and weak forces are all spin 1 particles (bosons), while the quarks and leptons are all spin 1/2 particles (fermions).
Supersymmetry (SUSY) solves this asymmetry in the spins of the particles by postulating  a new symmetry between bosons and fermions, which can only be realized in nature, if their exists for every particle with spin $j$ a supersymmetric partner with spin $[j \pm1/2$, thus doubling the particle content in the minimal extension of the SM, the so-called  "Minimal Supersymmetric Standard Model" (MSSM). The supersymmetric partners, usually depicted with a tilde on top (see Fig. \ref{f3}), have the same quantum numbers and interactions as the SM particles, but since the SUSY particles have not been observed at colliders in spite of intensive searches, the masses of the superpartners must be heavier than the SM particles. This breaks the (super)symmetry, but the mechanism of symmetry breaking is unknown. 
Here we simply summarize the reasons why Supersymmetry is so popular:

i)	SUSY is not plagued by the large quadratic corrections of the top mass to the Higgs mass, since the stop mass enters as a boson with an opposite sign in the amplitudes. Therefore no arbitrary low energy cut-off needs to be introduced in SUSY, but the theory is valid up to high energies.  

ii)	The electroweak symmetry breaking, which gives mass to the electroweak gauge bosons via the Higgs mechanism, occurs naturally via radiative corrections in SUSY. In contrast, this symmetry breaking needed to be introduced ad hoc in the SM.  Furthermore, this radiative electroweak breaking leads to the prediction of a top mass between 140 and 190 GeV and a Higgs mass below 135 GeV. These predictions were made before the discovery of these particles and turned out to be true: the top quark was discovered at a mass of 171 GeV and the Higgs boson at 125 GeV.

iii)	At high energies the gauge and Yukawa couplings unify at the GUT scale of $2\cdot10^{16}$ GeV for SUSY particles in the TeV range, thus paving the way for a Grand Unified Theory (GUT) of the strong and electroweak interactions, as we showed in a well-known paper \cite{Amaldi:1991cn}. In a GUT the three low energy forces of the SM with its symmetries based on the $SU(3)\otimes SU(2)_L\otimes U(1)$ are unified in a larger symmetry, like SU(5) or SO(10), which has the SM symmetries as subgroup. The larger GUT symmetry has additional gauge bosons which provide interactions between quarks and leptons, thus unifying them. These new gauge bosons get presumably mass during the symmetry breaking at the GUT scale, via mechanism similar to the Higgs mechanism giving mass to the electroweak bosons. During the symmetry breaking of the GUT symmetry into the SM symmetries all conditions for a baryon asymmetry, as spelled out by Sacharov \cite{Sakharov:1967dj}  (baryon- and lepton number violation; CP, C and P violation; non-equilibrium) are possible. In a GUT it is thus possible to explain why not all nuclei annihilated with anti-nuclei, so a few nuclei remained and formed the baryonic matter (including us) of the universe. The unification of quarks and leptons also explains why the electric charges of quarks and leptons are correlated, thus explaining why the hydrogen atom is perfectly neutral. Note that the Sakharov conditions are not obviously fulfilled in the SM, since e.g. the observed CP violation  is many orders of magnitude too small \cite{Rubakov:1996vz}.

iv)	SUSY predicts many new particles and the lightest supersymmetric particle (LSP) is expected to be stable, because of R-parity conservation. It turns out that in a large region of parameter space the LSP, usually called a neutralino, has exactly the properties of a WIMP. Its cross sections with normal matter have been reviewed in Ref. \cite{Jungman:1995df}. Especially, the  relic density of dark matter is obtained in Supersymmetry over a rather wide region of parameter space, which is sometimes called the ''WIMP miracle''.

As mentioned above, the masses of the superpartners are not known, so in principle all the masses and the mixing matrices are free parameters, which leads to a total of 105 free parameters. However, if one assumes that not only the gauge couplings unify, but also the soft mass breaking terms of the fermions and bosons have a common value at the GUT scale and obtains the Higgs mechanism via electroweak symmetry breaking (EWSB)  by radiative corrections, one is left with only 4 free parameters in the so-called Constrained Minimal Supersymmetric Model (CMSSM). An additional argument for unification is provided by the ratio of the masses of the b-quark and tau-leptons: in a larger GUT group, like SU(5), these fermions are in the same multiplet and if one assumes their Yukawa couplings are unified at the GUT scale, one obtains the correct mass ratio for these fermions at low energies. The four parameters of the CMSSM are: a common mass scale $m_0\  (m_{1/2})$ for spin 0 and spin 1/2, respectively, and two parameters in the Higgs sector: the trilinear coupling $A_0$ and $\tan\beta=v_2/v_1$, which is the ratio of the non-zero expectation values of the neutral Higgs components of the two Higgs doublets and the sign of the Higgs mixing parameter $\mu$, whose value is obtained from EWSB. These four parameters can be subjected to constraints from SUSY searches at accelerators, direct DM searches in deep underground experiments and cosmological constraints, i.e. the relic density, as studied  in great detail by many groups, see Ref. [15] and references therein. However, with the discovery of the Higgs mass at 125 GeV the CMSSM got slightly into difficulties: at Born level the Higgs mass is predicted to be below the Z$^0$ boson mass, i.e. below 91 GeV. To get a mass above the Z$^0$-mass is possible by radiative corrections, but these loop corrections are proportional to the logarithm of the stop mass, so stop masses in the multi-TeV range are needed in the CMSSM. But such large masses compromise the cancellation of the quadratic radiative corrections, which only works if top and stop masses are not too different. 

A solution to this problem is the Next-to-Minimal-Supersymmetric-Model (NMSSM), which has an extended Higgs sector, namely the NMSSM has a Higgs singlet in addition to the usual two doublets from the minimal models. Such a singlet is well motivated in many models (see Ref. \cite{Ellwanger:2009dp} for a review), but for the Higgs mass it has the advantage that the mixing with the singlet can lift the Higgs mass already at Born level above the Z$^0$-boson mass, so no heavy stop masses are needed to obtain a 125 GeV Higgs mass, see Refs. \cite{Ellwanger:2009dp,Beskidt:2013gia} and references therein. An additional spin-0 Higgs singlet in the NMSSM requires also the introduction of a spin 1/2 singlino, which mixes with the other gauginos and Higgsinos.  In a large region of the parameter space the LSP has a large singlino component. In this case the LSP is typically light and independent of the other SUSY masses in contrast to the CMSSM, where the LSP is typically a gaugino, which is related to the other SUSY masses. The differences between the CMSSM and NMSSM have important phenomenological consequences for direct DM searches, see Ref.  \cite{Beskidt:2017xsd} and references therein, especially the LSP is now singlino-like, i.e. it is largely the partner of the Higgs singlet. Therefore, its couplings to the SM particles mainly originates from the mixing with other Higgsino particles leading to dominant Higgs exchange in the diagrams of Fig. \ref{f2}, which automatically can explain the 10 orders of magnitude difference between the annihilation and scattering cross sections.
\begin{figure}[t]
\centering
\includegraphics[width=0.7\textwidth,clip]{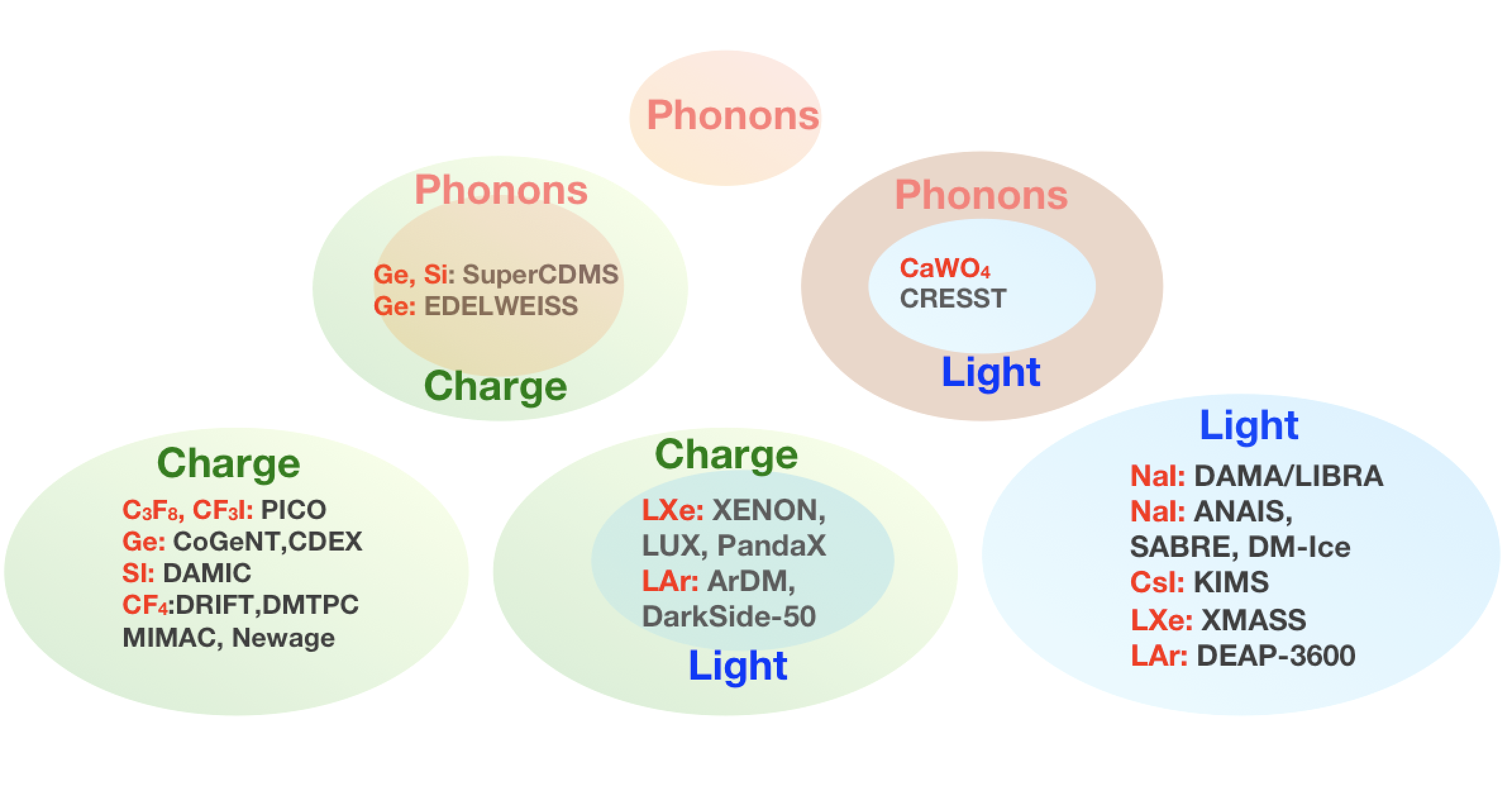}
\caption{Nuclear recoil signals used in various direct DM search experiments and the target materials (red).
}
\label{f4}       
\end{figure}

\section{DM searches and interpretations} 
\label{detection}
Fig. \ref{f2} summarized the basic DM Feynman diagrams for  direct scattering,  annihilation  and production at the LHC. All three have been discussed in detail at this conference \cite{morselli,fiorillo,rozen}. Although naively one would expect to be able to compare limits on the DM production at the LHC (right panel in Fig. \ref{f2}) with limits from direct DM searches (left panel), in practice this is difficult, because of the six orders of magnitude different energy scales involved: at the LHC the Higgs propagator couples to the proton by gluon fusion, while in the direct scattering experiments the Higgs propagator couples to the proton mainly by the interaction with the virtual s-quarks. Hence, the processes are different, which makes it difficult to compare the results.
Therefore,  people have considered to compare the processes in the framework of simplified models with arbitrary constant couplings \cite{Boveia:2016mrp,Buchmueller:2017qhf} and recent results have been presented at this conference \cite{rozen}. However, meaningful comparisons are only possible in realistic physical models, but in  realistic models, like the NMSSM model, the DM particle is largely a singlino and with the present luminosity no region of the allowed parameter space is excluded by the LHC  searches, so it will not be discussed here any more.


\subsection{Direct DM searches}

In these searches one looks for nuclear recoil signals from WIMPs hitting the nuclei in a target material, which can displace the nuclei from its lattice position. For a lattice displacement typically of few tens of eV energies are needed, while the momentum transfer by a WIMP in an elastic scattering on a nucleus is typically of the order of 1-100 keV, so a nuclear recoil is  displaced by many lattice spacings. These slow recoils can ionize or excite other nuclei and create phonons from the lattice vibrations.  For low momentum transfers the scattering happens coherently on the whole nucleus, so the cross section is proportional to the ''size'' of a nucleus, i.e. proportional to the nucleon number $A^2$. This cross section is ''spin-independent'' (SI), in which case the propagator is e.g. a spinless Higgs particle providing ''scalar'' interactions. Interactions by the exchange of a Z$^0$ boson are different for left- and right-handed particles. Since the spin is an axial vector, which changes sign, if watched in a mirror, or more physical, under the parity operator, thus leading to ''pseudo-scalar'' or ''axial vector'' interactions. Therefore, the unpaired spins $j$ of a nucleus are important and the ''spin dependent'' (SD) cross section is proportional to $j(j+1)$. The SD interactions occur at the quark level, so there is not the $A^2$ enhancement from the coherent scattering on all nuclei. 

The   observable signatures from nuclear recoils, like ionization, the heat (phonons) produced by it or scintillation light from excited nuclei,  are small and rare signals, so background signals from other particles entering the target material are a major concern. Therefore, the direct searches are performed in deep underground experiments to reduce the background from cosmic ray particles entering the detector. Further backgrounds are from radioactive elements present  in most materials. For example, a human body of 70 kg contains   0.0164 grams of radioactive potassium $^{40}$K leading to  4,300 disintegrations per second with 89\% of the integrations leading to an electron from $\beta$-decay with a maximum energy of 1.4 MeV. The remaining 11\% of the decays lead to gammay-rays with a maximum energy of 1.3 MeV. So a human being next to a target for DM searches would be a disaster, since typical DM interaction rates for the extremely small upper limit on the cross section of 10$^{-46}$ cm$^2$ are of the order of 5-10 events/tonne/year.

Methods to reduce the background are, as mentioned already, the use of deep 
 underground laboratories and the use of ultra-pure materials with low natural radioactivity. Furthermore, one can use detectors with position information, so the target can be divided into an inner region (fiducial volume) shielded by an outer region of target material and use at least two different signals, by which one can discriminate between nuclear recoils and electromagnetic interactions from electrons and gamma-rays from radioactive materials. The latter particles tend to transfer a large fraction of their energy to the bound electons in  the target material, simply because they have the same mass (just like two billiard balls can transfer a large fraction of the energy from one to another). Note that electrons hitting a nucleus of the target transfer little energy, just as a tennis ball hitting the earth does not move the earth very much. Hence, electrons and gamma-rays easily produce a large ionization signal, but hardly produce nuclear recoil signals, like scintillation light, ionization or  heat from a moving nucleus. However, neutrons,  produced  by nearby interacting cosmic rays, can produce nuclear recoils. 
 
 Many detectors have been used using all possible materials and signals or combinations of them, as summarized in Fig. \ref{f4} and many reviews exist, see e.g. \cite{Undagoitia:2015gya,Schumann:2015wfa,Patrignani:2016xqp,Liu:2017drf} and references therein. 
  To measure the tiny temperature increase from nuclear recoils the detectors are cooled down by dilution refrigerators to temperatures into the mK range, which reduces the heat capacity, so small energy deposits lead to large temperature variations in  bolometers or to large electric resistance changes in superconductors operated at the edge of the transition to superconductivity. 
  The flux of WIMPs is proportional to the relative velocity of the detector and the WIMPs (just like the rate of rain drops hitting a person  increases if he starts running). This leads to an annual modulation of the event rate with a maximum in June, when  the Earth has  the maximum velocity component in the direction of the movement of the Sun, while a minimum is expected in December \cite{Freese:2012xd}.  This annual modulation was measured over 14 years by the DAMA and DAMA/LIBRA experiments in the national underground laboratory in Italy under the Gran Sasso mountains (Laboratori Nazionali del Gran Sasso) \cite{Bernabei:2013xsa}. They find with a significance of more than 9$\sigma$ two allowed regions: a  WIMP cross section of about 10$^{-40}$ cm$^{2}$ (10$^{-41}$ cm$^{2}$) for WIMPs masses around 10 (50)  GeV \cite{Patrignani:2016xqp}. These regions are excluded by other experiments by a large margin, so  it is  hard to imagine that the different target materials can account for it, since this would require strong isospin violating WIMP particles leading to largely different cross sections for neutrons and protons. Therefore, the origin of this disagreement is not understood. It may have to do with the large background in the experiment, which may also have an annual modulation, but it is surprising that the background has the same phase of the modulation.  
   The KIMS experiment  does not confirm the annual modulation signal from DAMA/LIBRA \cite{Kim:2012wuh}, but they use  CsI(TI) instead of NaI(TI) crystals. Also  the DM-ICE collaboration, which employs  NaI(TI) crystals under the ice of the Southpole, does not observe modulation, but their sensitivity is not yet  enough to contradict the DAMA/LIBRA results \cite{deSouza:2016fxg}. Future experiments   will operate with NaI(TI) crystals, but the trouble is that the crystal growing company for DAMA (Saint Gobain Crystals) does not deliver the radiopure crystals to other experiments because of a confidential restrictrion with DAMA. Another puzzling aspect of the DAMA/LIBRA results is the fact that the modulation amplitude decreases with time, see Ref. \cite{Patrignani:2016xqp}. Here also the future experiments (ANAIS \cite{Coarasa:2017aol}, DM-ICE, SABRE \cite{Froborg:2016ova,Tomei:2017rkg} , KIMS \cite{Kim:2012wuh}) using NaI(TI) crystals hopefully will shed light on the origin of the DAMA/LIBRA modulation signal. The SABRE collaboration plans to use an active veto shield and to operate twin detectors on the northern and southern hemisphere. If the origin has to do with seasonal variations the different hemispheres should have a different phase, while a DM origin would have the same phase in both hemispheres.
 \begin{figure}[t]
\centering
\includegraphics[width=0.45\textwidth,clip]{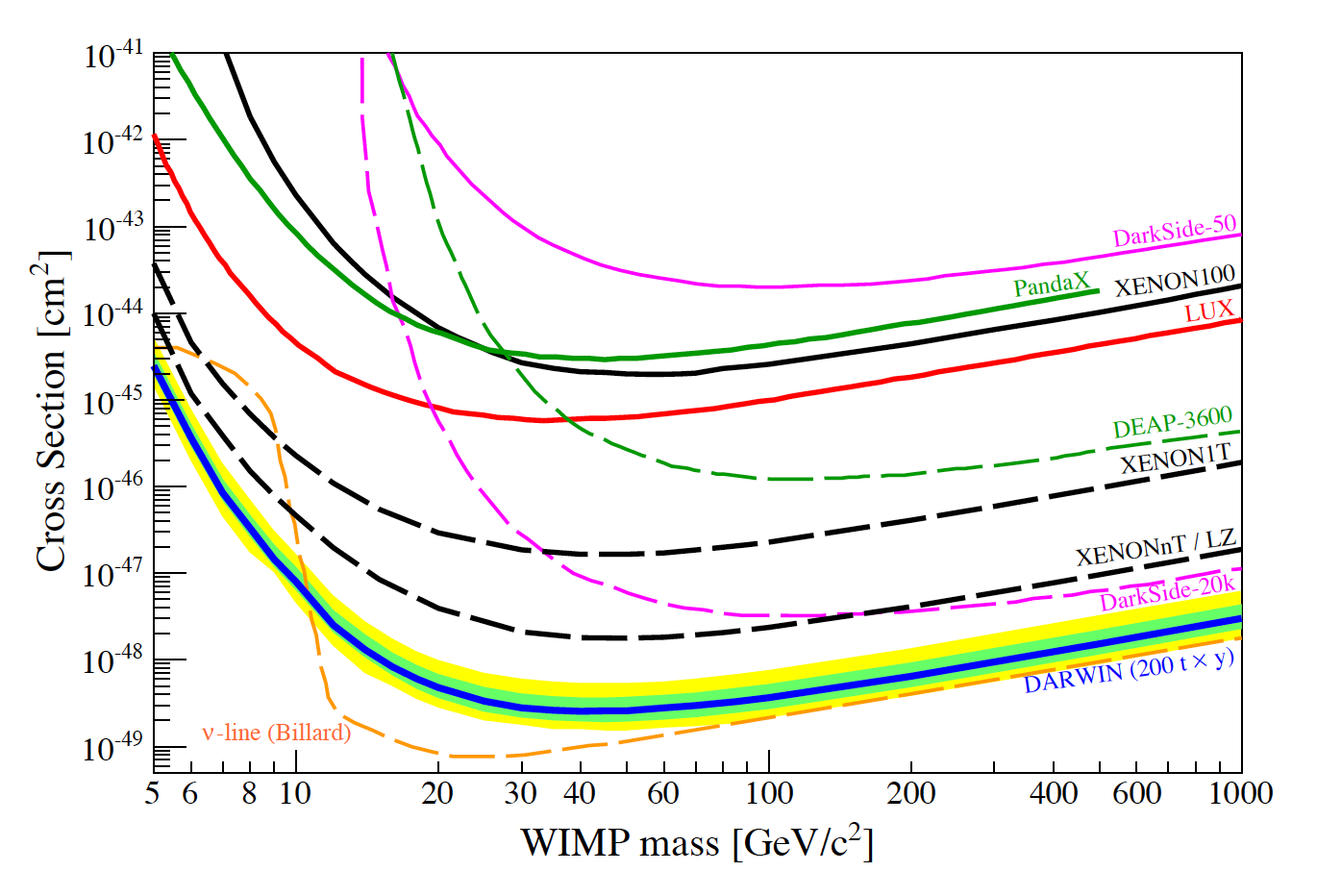}
\includegraphics[width=0.45\textwidth,clip]{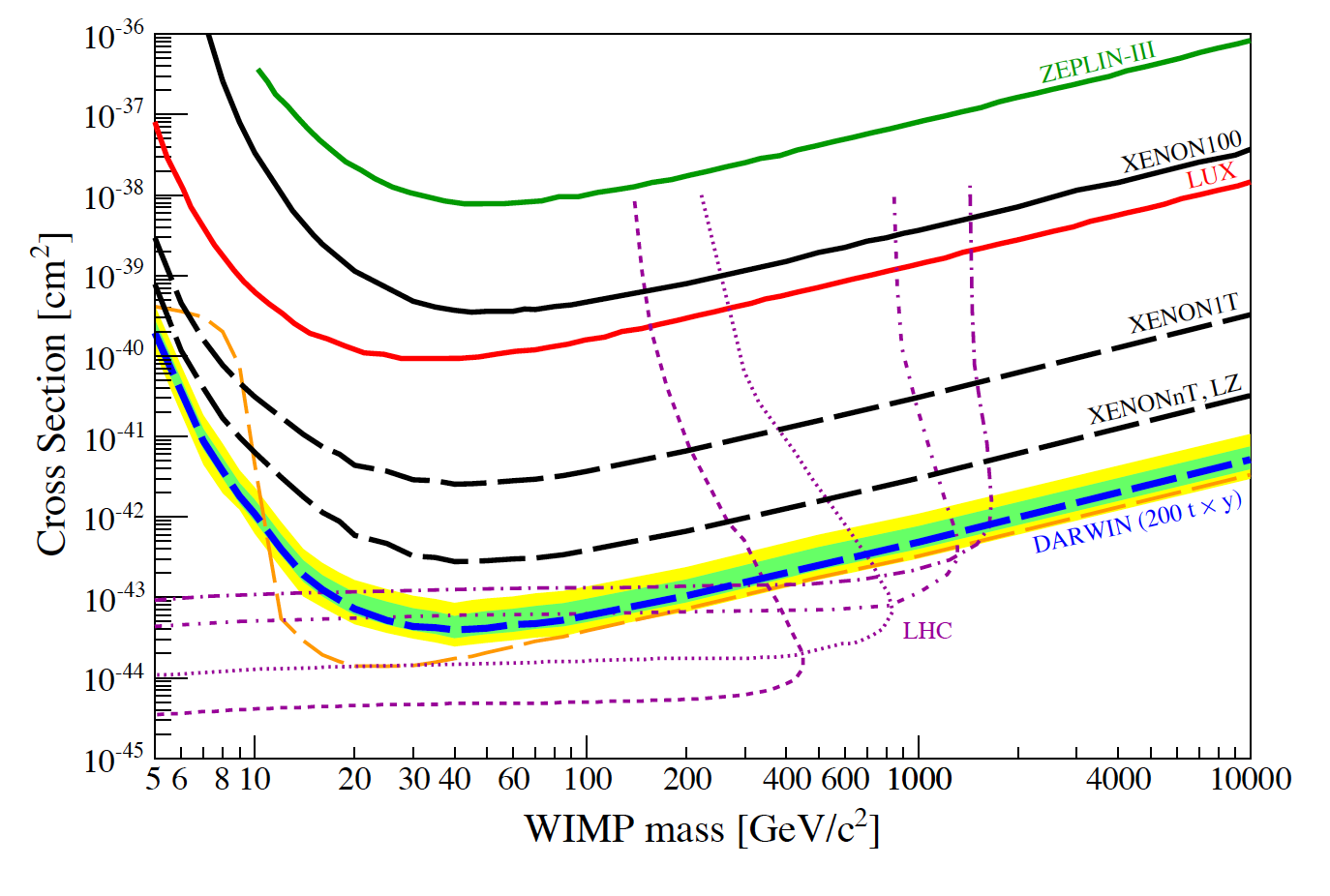}
\caption{WIMP cross section  limits from present  direct DM search experiments for SI- (left) and SD-interactions (right) as well as the expected sensitivity for future experiments (dashed lines). From Ref. \cite{Aalbers:2016jon}.
}
\label{f5}       
\end{figure}
\begin{figure}[t]
\centering
\includegraphics[width=0.8\textwidth,clip]{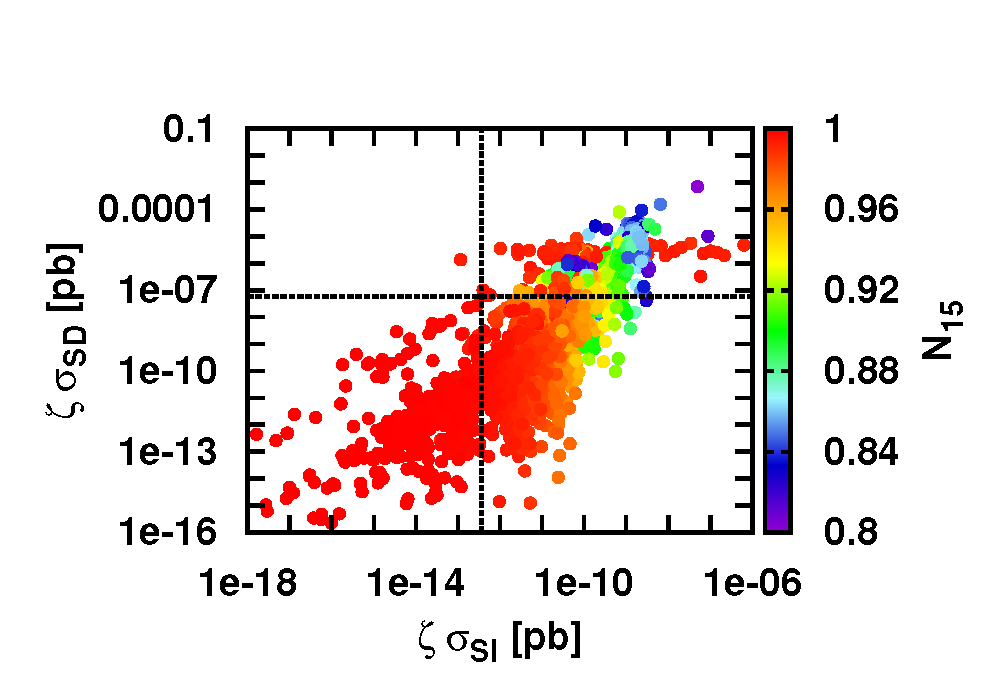}
\caption{SD versus SI WIMP scattering cross sections in the NMSSM. The colour coding represents the singlino contribution to the LSP. For almost pure singlinos the cross sections are well below the neutrino floor for atmospheric neutrinos, as indicated by the  vertical and horizontal dashed lines. From Ref. \cite{Beskidt:2017xsd}.
}
\label{f6}       
\end{figure}

 Liquid noble gases reach nowadays the highest sensitivity, since they can have large target masses with high purity. 
As an example of a noble gas detector we discuss a XENON detector consisting of a dual phase time projection chamber (TPC), which consists of  a liquid xenon target with a gaseous phase on top. The principle is as follows. A WIMP scattering on a Xenon nucleus 
 produces scintillation and ionization. The prompt scintillation light produces 178 nm ultraviolet photons, which produces the S1 signal in the photomultipliers (PMTs) on the top and bottom  of the Xenon.  The electrons from the ionization  drift to the top of the liquid phase by the electric field and enter the gas phase, where the strong  electric field accelerates the electrons to energies high enough to create another scintillation signal S2 in the PMTs, which is delayed by the drift time in the liquid phase. So from the time difference between S1 and S2 the vertical coordinate can be determined, while the position of the PMTs determines the coordinates in the horizontal plane. This 3-D determination of the events inside the detector allows to determine the background from the outside, which rapidly decreases towards the fiducial volume in the centre by the self-shielding of the outer layers. Such a pixelation allows  a rigorous control of the background, which is the main game in the search   for very rare events.
 Recent and future limits on the direct DM scattering cross sections have been summarized in Fig. \ref{f5}. Future experiments are expected to cover the region down to the "neutrino floor" \cite{Billard:2013qya}, as indicated in Fig. \ref{f5} by the lowest dashed line. The large background at low WIMP masses originates from solar neutrinos, while above 10 GeV the background is dominated by scattering from the more energetic neutrinos produced by diffuse supernova neutrinos or cosmic rays in the atmosphere.

 Below this floor detection is expected to be notoriously difficult, since one can only do counting experiments and even with a 40 tonnes Xenon detector the event rate is only 45 events per year  for a cross section of $2\cdot10^{-47}$ cm$^2$ \cite{baudis}. However, if the DM is an almost pure singlino, it hardly couples to visible matter and the expected cross sections can be as low as $2\cdot10^{-52}$ cm$^2$ for SI scattering \cite{Beskidt:2017xsd}. This is well below the neutrino floor, which is indicated in Fig. \ref{f6} by the vertical and horizontal dashed lines. Detectors  with directional information may help to suppress background, since WIMP signals and the neutrino background have a different distribution of the recoil angle, since the first one shows a dipole  feature, peaking around the direction of the Solar motion, while the neutrino background is more isotropic \cite{Mayet:2016zxu}.

\subsection{Indirect DM detection}
\label{indirect} 
As mentioned before,  the annihilation cross section can be determined from the Hubble constant and its value is given in the central panel of Fig. \ref{f2}, which is ten orders of magnitude higher than the upper limit on the scattering cross section, given in the left panel. This is most easily be explained, if the exchanged particle is a Higgs boson, since this strongly suppresses the scattering on the light quarks of a nucleon. This also implies that the annihilation is predominantly into heavy quarks and leptons (90\% into b-quark pairs, 10\% into tau-lepton pairs). The dominance of the b-quarks over tau-leptons is a combination of the higher b-quark mass  and the fact that the b-quark comes in three colours.  Although the annihilation stops after freeze-out (about 1 ns after the Big Bang), it can restart much later in the galaxies, where the DM concentrates, so particles are close enough to annihilate leading to a flux proportional to the density of DM particles squared. N-body simulations indicate that the DM density  follows an NFW-profile, which falls off as function of radius like 1/R$^2$ above a certain scale and  like 1/R near the galactic centre. So the highest annihilation rate is expected to be in the centre. 
With the annihilation into b-quark pairs the decays will include among others gamma-rays from $\pi^0$ decays, positrons and electrons, antiprotons and protons, antineutrinos and neutrinos, antideuterons and deuterons. The rate and spectra for these particles are reasonably well known from the LEP experiments, which studied the annihilation of electron-positron pairs up to centre-of-mass energies of 214 GeV and the final state in annihilation is independent of the initial state. Given the preponderance of matter in our Galaxy it is easier to search for antimatter particles from  DM annihilation.  Antimatter and gamma-rays can also be produced by nuclear interactions of cosmic rays with the gas of the Galaxy. However, the energy spectrum from the two sources are different: the  energy spectra from annihilation are harder  with a sharp cut-off at  the WIMP mass and the spatial production of
the DM annihilation follows an NFW-like profile, thus being spherically distributed around the   Galactic centre on a scale an order of magnitude broader than the  visible matter and sharply peaking in the centre. Unfortunately, the charged particles are not pointing back to the production site because of the magnetic fields in the Galaxy, so one cannot use the spatial information, only the spectral energy information. Gamma-rays, on the other hand, point back to the source, so one can check if the spatial information is consistent with an NFW-like profile.  Given the large absorption of the atmosphere the indirect DM searches are best performed with satellites outside the atmosphere, although air showers can also be used to detect cosmic rays. However, here the background from misidentified hadrons is large and the energy thresholds are in general large, so  WIMPs below 100 GeV will be hard to detect, see e.g. the H.E.S.S results \cite{Abdallah:2016ygi} or the prospects of the future CTA project \cite{Acharya:2017ttl} for the detection of gamma-rays from the Galactic centre or neutrino results from ICECUBE \cite{Aartsen:2017ulx} and Super-Kamiokande \cite{Desai:2004pq}. The neutrino experiments can also look for an enhanced DM density in the Sun
by gravitational trapping.
\begin{figure}[t]
\centering
\includegraphics[width=0.7\textwidth,clip]{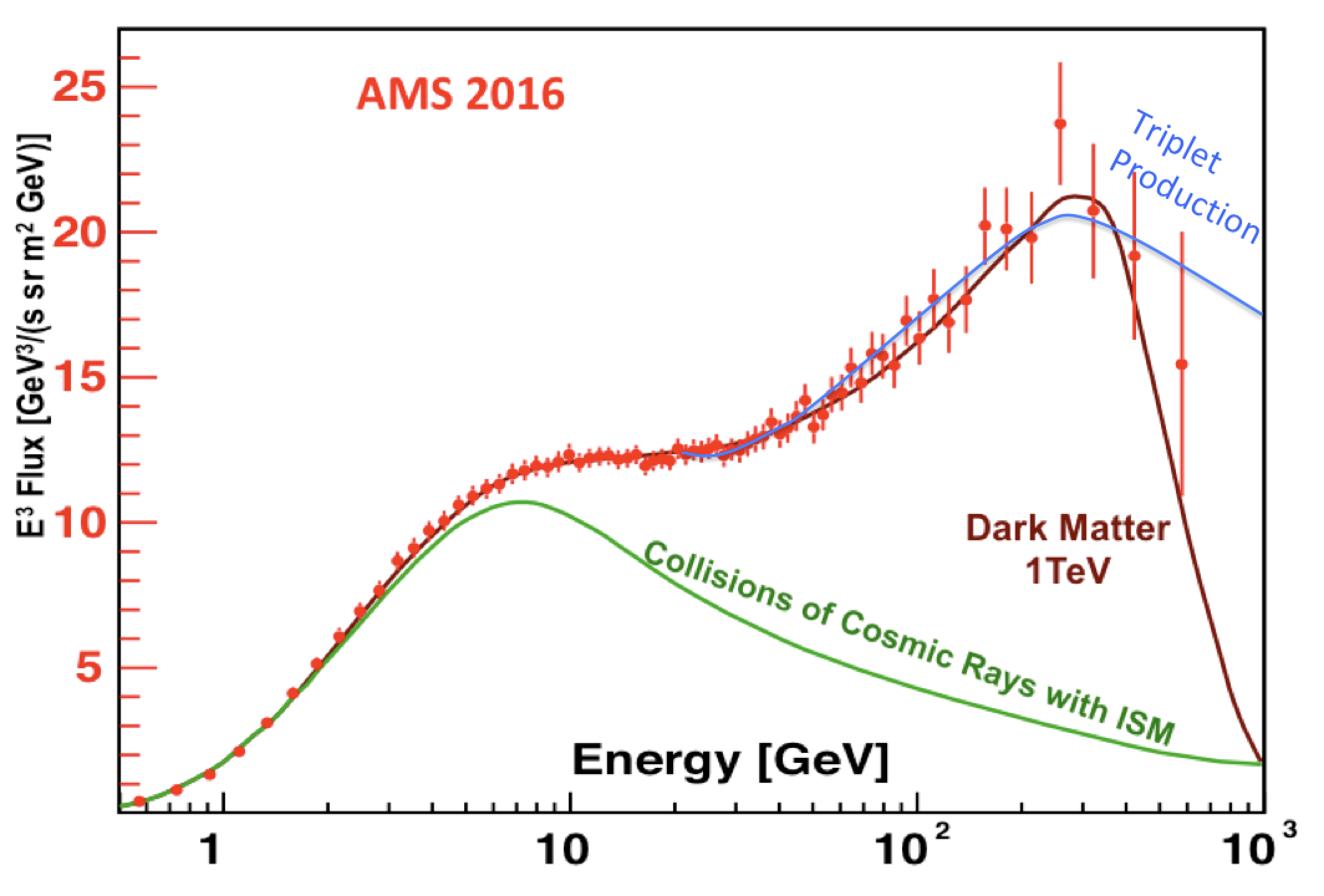}
\caption{The positron flux from AMS-02  in comparison with background, annihilation from a 1 TeV WIMP \cite{AMS-positron} and triplet production \cite{biermann}.
}
\label{f7}       
\end{figure}
Several signals have been hotly debated as a sign for a signal from DM annihilation, like 1) the high energy bump in the positron flux, as observed by  AMS-02; and 2) the FERMI-GeV excess of diffuse gamma-rays. Other searches have given limits, like the neutrinos searches from Superkamiokande and Icecube  and gamma-rays from dwarf-galaxies using Fermi-data \cite{Patrignani:2016xqp, morselli}. The results have all been carefully documented in the Particle Data Book  \cite{Patrignani:2016xqp} and will not be discussed here. We shortly discuss the two signals mentioned above as examples of the difficulties in excluding other astrophysical explanations.
\subsubsection{The positron excess}
The locally measured positron  spectra from AMS-02 is shown in Fig. \ref{f7} \cite{AMS-positron}. The positron flux  increase above 30 GeV and falls above 250 GeV. This could be explained by a DM annihilation with a WIMP mass around 1 TeV, but pulsars with a cut-off might explain the results as well, see e.g. Ref. \cite{Hooper:2017gtd} and references therein for a recent evaluation using the high energy gamma-ray flux from nearby pulsars observed by HAWC. However, the HAWC collaboration disputes this interpretation and thinks other processes are more probable \cite{Abeysekara:2017old}. The difficulties are connected with the escape, flux cut-off and propagation from the positrons originating from pulsars.   More mundane explanations with less free parameters are the triplet production,  in which a photon is converted in the electric field of an electron, which then looses energy. This is one of the well known energy loss processes  studied long ago \cite{Haug:1975bk,Haug:1981za,2004A&A...416..437H}. If the electron spectrum inside sources has the expected 1/E$^2$ spectrum, as observed for the source cosmic rays \cite{deBoer:2014bra,deBoer:2017sxb}, one obtains a shape, which well describes the increase in the positron flux, as indicated by the solid blue line in Fig. \ref{f7}. Also from the enhanced CR density and high photon density in sources the order of magnitude of the triplet production is acceptable \cite{biermann}. Fig. \ref{f7} shows also the expected signal from a 1 TeV WIMP by the black curve. Note that the endpoint of the DM curve can be increased by a higher WIMP mass, while the endpoint of the triplet production can be reduced by a lower cut-off of the electron spectrum, which was here chosen to be 30 TeV.
So both solutions can be tuned to describe the positron excess. Note that only the shape can be compared with the data, since the absolute fluxes  depend on the propagation from the source to the earth, which has large uncertainties. Alternatively, one can try to explain the spectral shape by propagation effects \cite{Cowsik:2013woa}.

\begin{figure}[]
\centering
\includegraphics[width=0.4\textwidth,clip]{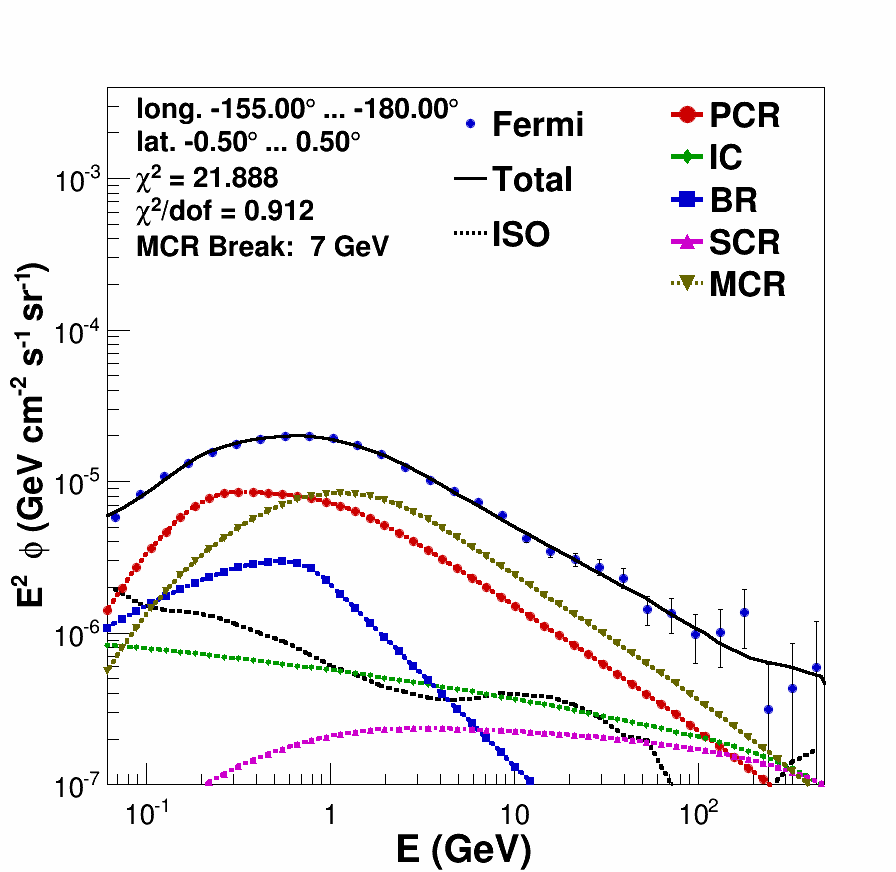}
\includegraphics[width=0.4\textwidth,clip]{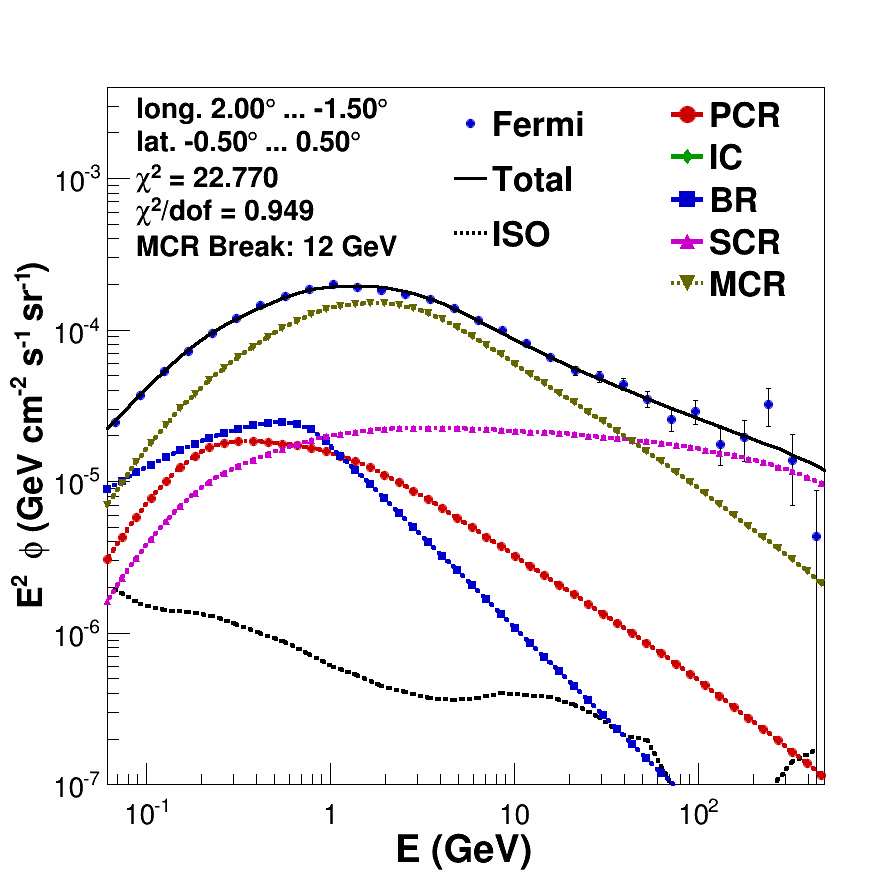}
\caption{Energy template fits to the Fermi diffuse gamma-ray data away from the Galactic Centre (left) and towards the Galactic Centre (right). The shift of the maximum from 0.7 to 2 GeV can be explained by the presence of the molecular cloud complex in the Galactic Centre, in which the emissivity at low energies is suppressed by energy losses and/or magnetic cut-offs, as shown by the MCR template. The hard spectrum from source cosmic rays (SCR template) is responsible for the high energy tail in the data, which is the origin of the so-called gamma-ray gradient problem \cite{Yang:2016jda}: the spectral hardening from the Galactic anti-centre to the Galactic Centre. The morphology of the SCR template follows closely the morphology of the $^{26}$Al line, which is a tracer of sources \cite{Prantzos1996}, thus proving that the spectral hardening is well described by the source cosmic rays \cite{deBoer:2014bra,deBoer:2017sxb}.
}
\label{f8}       
\end{figure}
\begin{figure}[]
\centering
\includegraphics[width=0.65\textwidth,height=0.6\textwidth,clip]{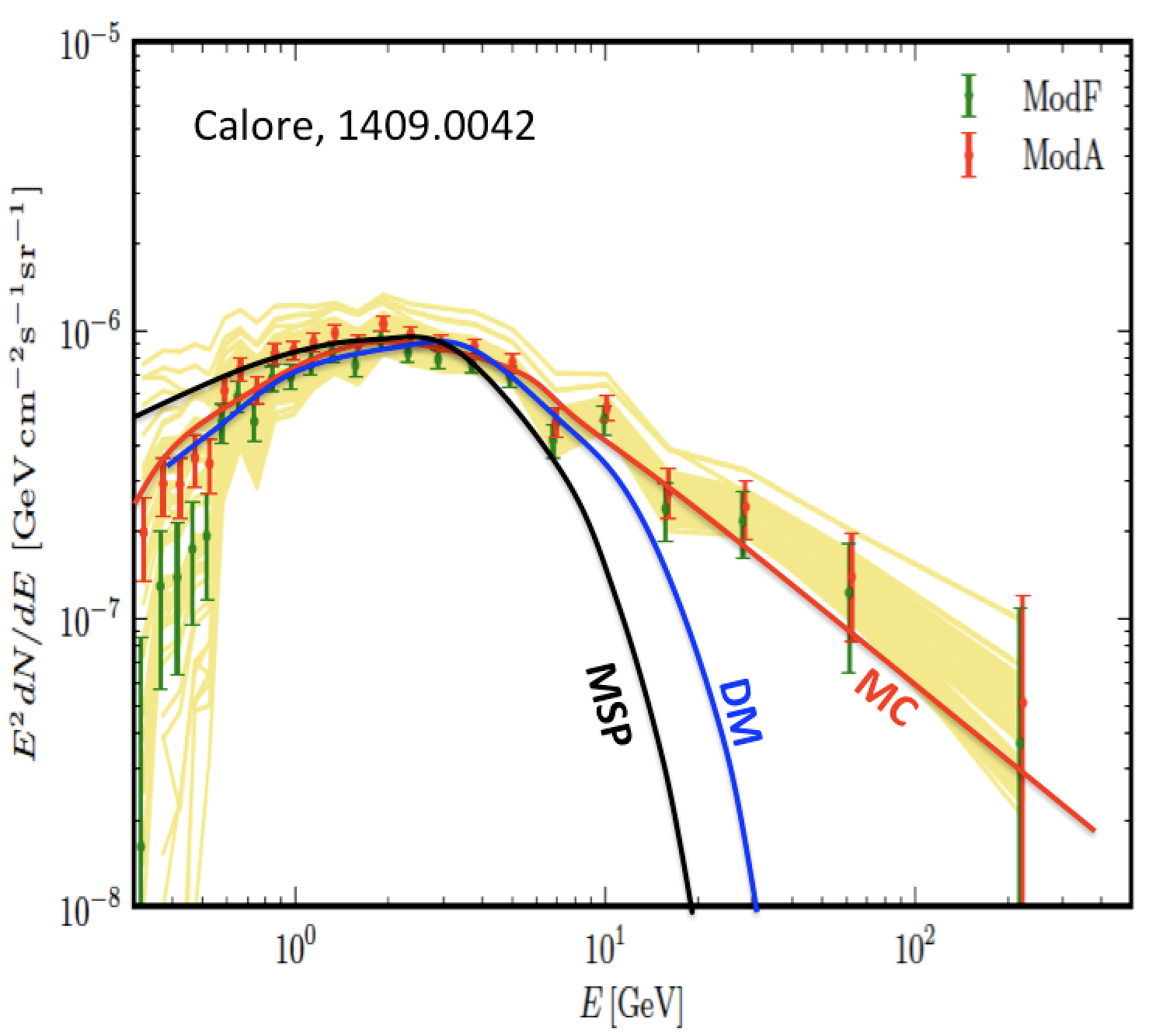}
\caption[]{The spectral shape of the excess using the Galprop propagation code to determine the background (data points from Ref. \cite{Calore:2014xka}) in comparison with spectral templates for molecular clouds (MC), dark matter (DM) (both from Ref. \cite{deBoer:2017sxb}) and  millisecond pulsars (MSP) (from Fig. 5 in Ref. \cite{Fermi-LAT:2017yoi}).
}

\label{f9}
\end{figure}
\subsubsection{The Fermi-GeV excess}
The flux of diffuse gamma-rays in the GeV regime shows an excess above the expectations from diffuse emission models, for which usually the three main backgrounds from (1) gamma-ray production in nuclear reactions of nuclear cosmic rays with the gas in the Galaxy (from $\pi^0$-production and decay) and from electromagnetic interactions between leptonic cosmic rays with the gas (2) (Bremsstrahlung (BR)) and the photons of the interstellar radiation field (ISRF) (3) (inverse Compton (IC) scattering). The spatial distribution of the diffuse gamma-ray sky from these backgrounds can be estimated from   the spatial distributions of the target materials (ISRF for IC, and gas for BR + $\pi^0$-production). 

Unfortunately, these backgrounds  poorly describe the gamma-ray sky maps and at least two prominent excesses were observed: the Fermi bubbles, observed as a giant excess below and above the disk extending up to latitudes of 50$^\circ$ (=10 kpc) above the disk and the  so-called GeV excess, observed as a shift  of the maximum of the gamma-ray spectrum from 0.7 GeV for the standard background expectation to up to 2 GeV. The shift is most prominent towards the Galactic centre, as shown  in  Fig. \ref{f8}, where we plotted the spectra towards the centre in comparison with the opposite direction and used the trick to plot the spectra as a ratio to a power law with a spectral index of -2 in order to see better the deviations from a power law.
\begin{figure}[]
\centering
\includegraphics[width=0.42\textwidth,clip]{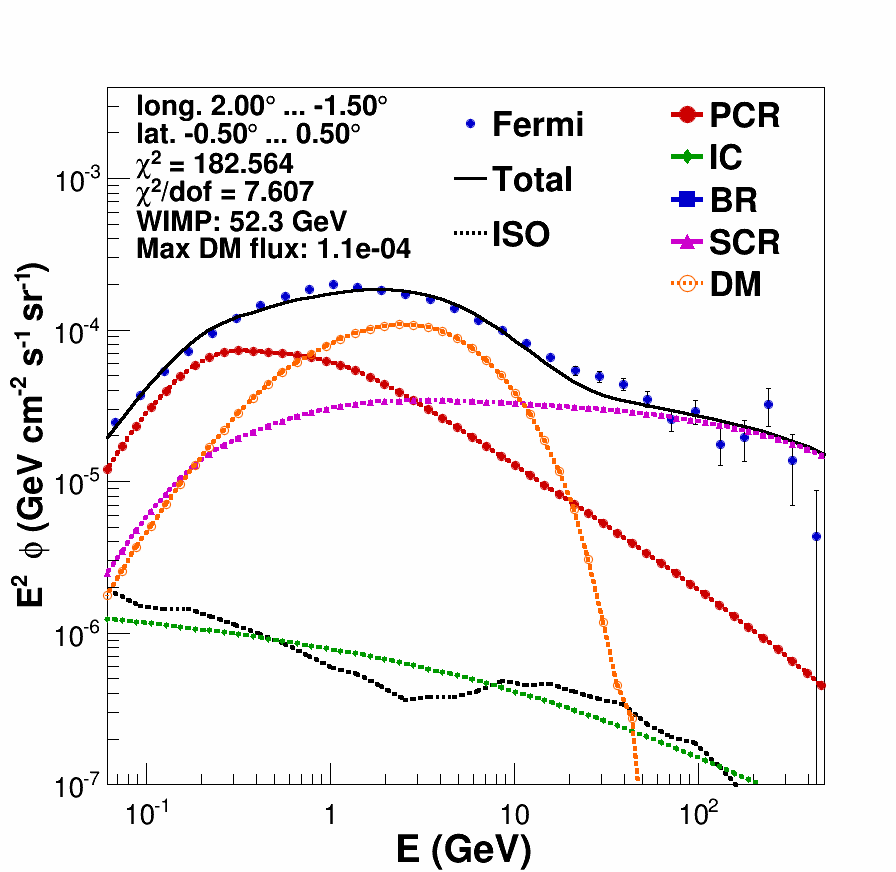}
\includegraphics[width=0.4\textwidth,clip]{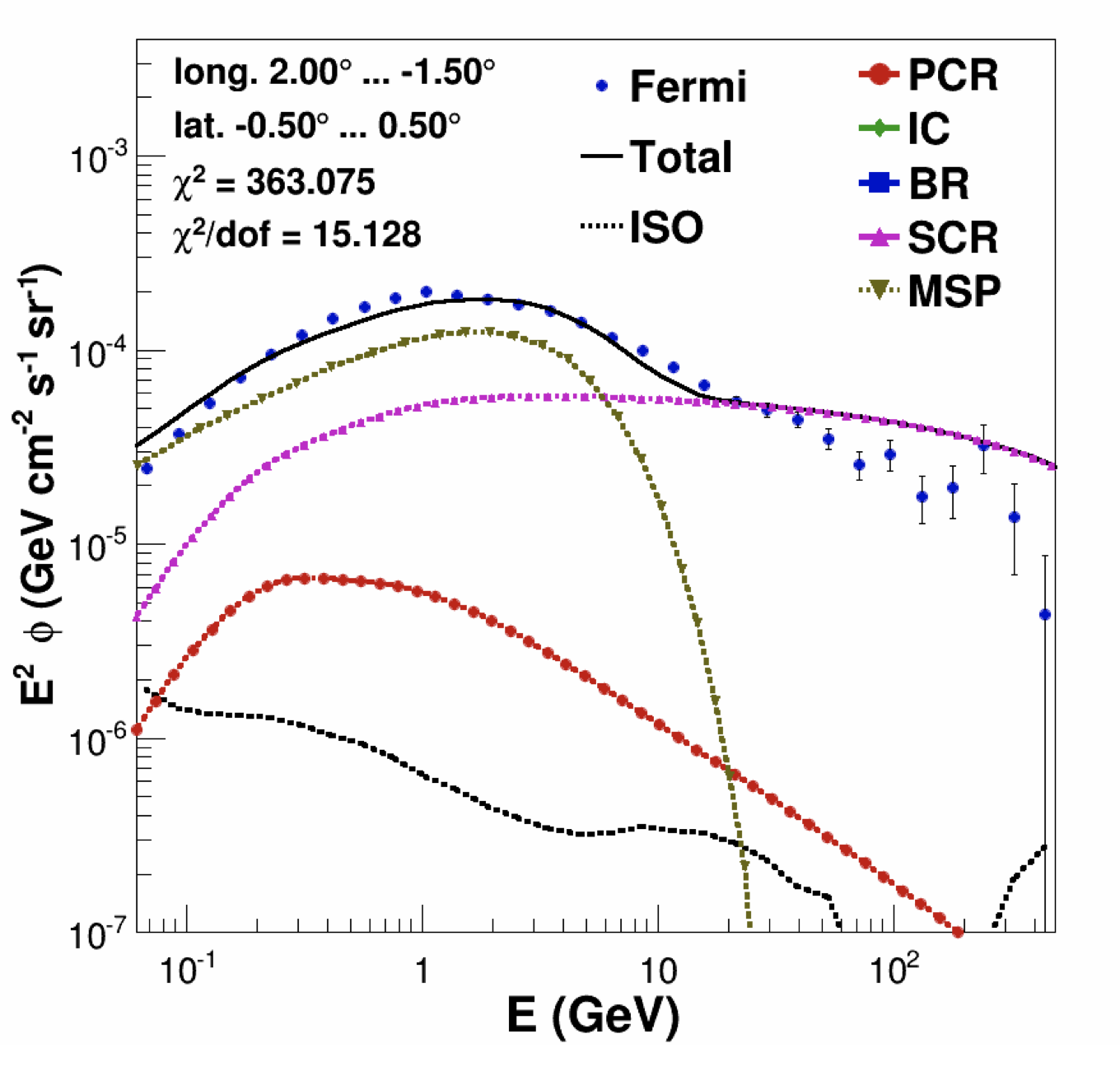}
\caption[]{As in Fig. \ref{f8}, but with the MCR template replaced by a DM template for a 52 GeV WIMP (left) or the MSP template (right) using the averaged energy spectrum from the observed millisecond pulsars, as given by the dashed line in the right panel of Fig. 5 of Ref. \cite{Fermi-LAT:2017yoi}. Clearly, the fit with the MCR template  in Fig. \ref{f8} yields much better $\chi^2$ values, as indicated in the figures.
}
\label{f10}[]       
\end{figure}
\begin{figure}[]
\centering
\includegraphics[width=0.4\textwidth,clip]{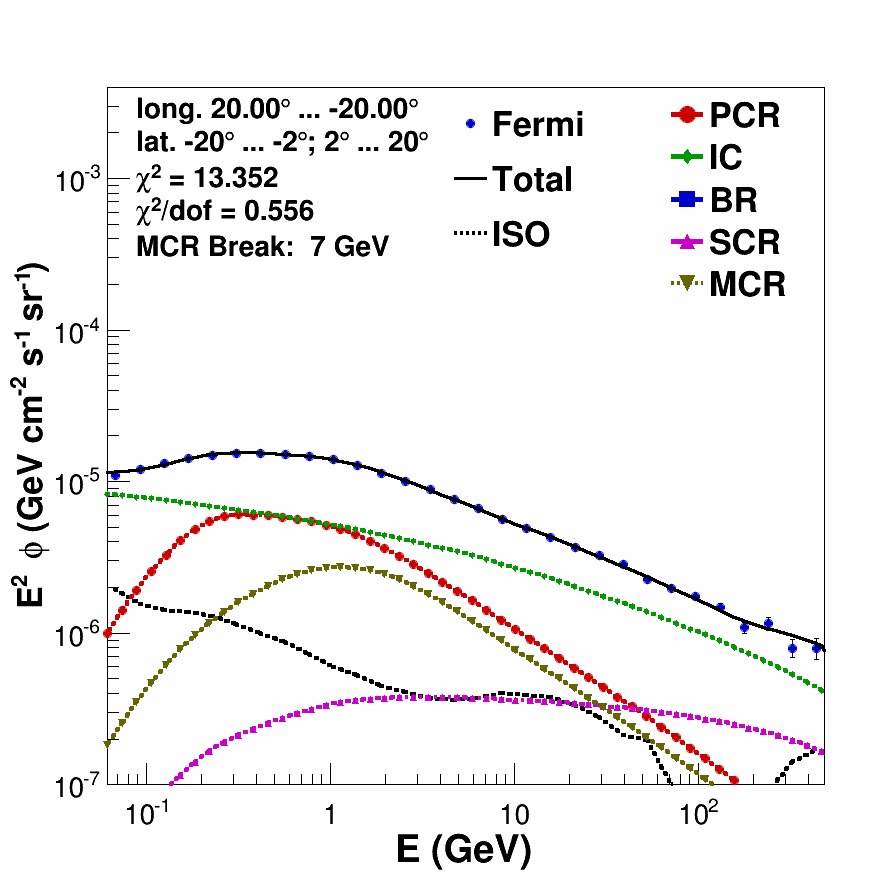}
\includegraphics[width=0.4\textwidth,clip]{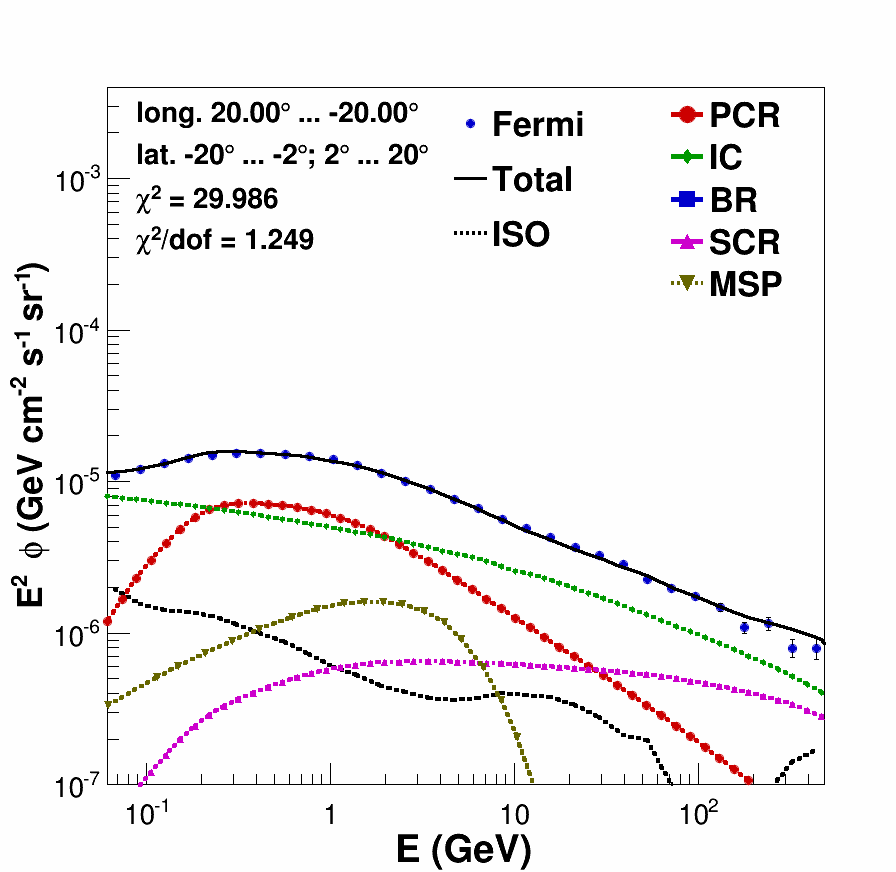}
\caption[]{As in Fig. \ref{f10}, but now for the halo region and comparing the molecular cloud MCR template (left) with the millisecond pulsars MSP template (right). Here the contributions from MCR and MSP are subdominant, so acceptable fits can be obtained with fluxes compatible with previous publications. 
}
\label{f11}      
\end{figure}
\begin{figure}[]
\centering
\includegraphics[width=0.7\textwidth,height=0.6\textwidth,clip]{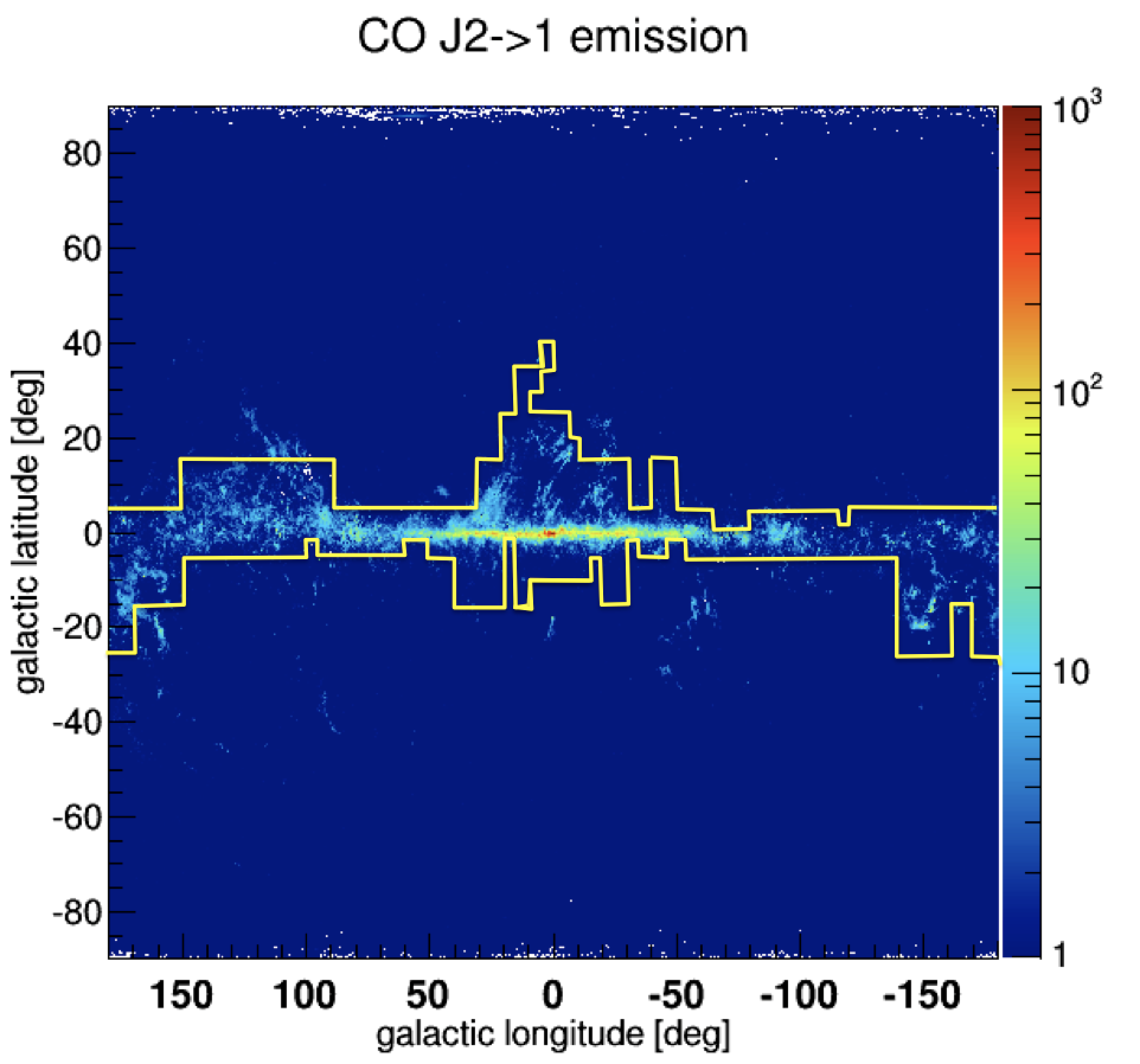}
\caption[]{Sky map of the CO rotation line as measured with the Planck satellite \cite{Planck}. The yellow (light) contour   is the sky map from the GeV excess, if fitted with the DM template from the left panel of Fig. \ref{f10}.
}
\label{f12}
\end{figure}
 The excess can be explained by the contribution of a new source with a spectrum peaking at 2 GeV. Three sources have been proposed: a dark matter (DM) annihilation signal  (from Ref. \cite{Daylan:2014rsa} and references therein), a signal from millisecond pulsars (MSPs) (from Ref. \cite{Bartels:2015aea} and references therein), and a signal from molecular clouds (MCs) (from Ref. \cite{deBoer:2017sxb} and references therein). All three have spectra peaking at 2 GeV, but slightly different shapes of the spectrum, as shown in Fig. \ref{f9}.
 Here the data are determined as an excess above the predicted background from a propagation model, which does not describe the data in the disk, so only the excess in the halo was determined (|b|>2$^\circ$) \cite{Calore:2014xka}. 
  One observes that the spectral shape from MCs, as determined from the spectrum of the Central Molecular Zone (CMZ), a  massive molecular cloud complex in the Galactic centre, describes the high energy tail  better than the spectra from DM and MSPs, for which the spectra are limited by cut-offs. There is no cut-off for molecular clouds, since here the high energy tail is determined by cosmic rays interacting with the gas and indeed, the spectral index of the excess at high energies is close to the spectral index of high energy protons.  
  To circumvent the use of poorly fitting propagation models for the background determination one can invoke spectral template fits, in which the spectrum in a certain sky direction is fitted as a linear combination of all physical processes, as discussed in Ref. \cite{deBoer:2017sxb}. In this case one determines the backgrounds from the fit instead of using external models and a missing contribution from a new physical process will shown up as a bad fit, while the region of the bad fit will determine the morphology of the new process. As expected already from the poor performance of propagation models one needs additional physical processes in addition to the  standard backgrounds mentioned above. As it turns out, including the predicted and observed $\pi^0$-production in the shocked gas from freshly accelerated protons inside sources (Source Cosmic Rays, SCR) \cite{deBoer:2014bra} and the enhanced energy losses and/or magnetic cut-offs for cosmic rays inside  molecular clouds  (MCR) provides excellent fits over the whole gamma-ray sky \cite{deBoer:2017sxb}. The correctness of including the MCR and SCR templates was proven by the fact that the fit needed MCR  templates only in directions of molecular clouds, as traced by the rotation lines of the CO molecules, while the SCR template was only needed in the direction of sources, as traced by the $^{26}$ Al line. The MCR template describes the excess by shifting the maximum of the gamma-ray spectrum to 2 GeV, while the SCR templates describes the high energy gamma-ray tail, as observed in sources and in the Fermi Bubbles, possibly an advective outflow of gas by the pressure of SCRs \cite{deBoer:2014bra}. Examples of template fits describing the shift of the spectrum from 0.7 GeV in the direction opposite to the GC and 2 GeV towards the GC are shown in Fig. \ref{f8}. A strong MCR component (= large shift) usually coincides with a strong SCR component (= large tail), since both are simultaneously present in molecular clouds. This strong correlation is apparent from the right panel in Fig. \ref{f8}. The hard SCR spectrum corresponds to a proton spectrum with a spectral index of  -2.1, as expected from diffuse shock wave acceleration. The data correspond to  7.5 years of Fermi Pass 8 diffuse gamma-ray data with the templates determined in a data-driven way, as described in Ref. \cite{deBoer:2017sxb}. The fit with the spectral shapes from DM and MSPs  are shown in Fig. \ref{f10}, which clearly yield a poorer value of the $\chi^2/d.o.f$, which was used as test statistic and is indicated in the figures.  
 
 The fit in the region of the CMZ in the right panel of Fig. \ref{f8} is dominated by the signal from cosmic rays interacting in the molecular cloud (MCR template), which is not surprising, since the CMZ harbors 5\% of the total molecular gas inside the tiny solid angle of lxb=4$^\circ$x1$^\circ$. Some people think this dominance must mean that it is a ''different thing'' in comparison with the GeV excess, but the excess was discovered by excluding the disk in the analysis and in the halo the flux is almost an order of magnitude lower, as shown in Fig. \ref{f11}. Here the excess is subdominant and acceptable fits can be obtained for all three proposed explanations of the excess. Only from the Galactic centre it is easy to distinguish between the proposed origins of the excess, as is obvious from the fits in Figs. \ref{f8} and \ref{f10}.  If the molecular clouds are responsible for the excess,  it should be not only located in the Galactic centre, but follows the distribution of MCs, which are distributed in the disk, as is known from the CO maps from the Planck satellite \cite{ThePlanck:2013dge}. In contrast, DM should peak in the Galactic centre with a steep fall-off in all directions. This is not the case, as demonstrated in Fig. \ref{f12}. Here  a fit with the DM template is overlaid with the sky map of the CO distribution. One observes that the excess obtained from a fit with DM follows  the CO profile. The fall-off in latitude  of the CO profile in the Galactic centre is steep (see Fig. \ref{f12}) and happens accidentally to be close to an expected DM profile, like an NFW profile. The reason for the sharp fall-off is the rapidly decreasing column density along the line-of-sight, which crosses a shorter fraction of the disk at increasing latitude.

In summary, if one considers a limited field-of-view centred within 20$^\circ$   around the Galactic centre and applies cuts on the energy range and/or excludes low latitudes,  cuts typically applied in the early analysis of the excess, all three hypotheses for the excess (dark matter, millisecond pulsars, molecular clouds) can describe the excess. However, if one considers the whole gamma-ray sky and includes gamma-ray energies up to 100 GeV one finds that the MC hypothesis is  preferred over the other hypotheses  for several reasons: i) The MC hypothesis provides significantly better fits; ii) The morphology of the  ''GeV-excess'' follows the morphology of the CO-maps, a tracer of MCs, i.e. there exists a strong  ''GeV-excess'' in the  Galactic disk also at large longitudes; iii) The  massive CMZ  with a rectangular field-of-view of $l \, \times \, b \; \simeq \; 3.5^{\circ} \, \times \, 1^{\circ}$ shows the maximum of   the diffuse gamma-ray spectrum at 2 GeV, i.e. the ``GeV-excess'', already in the raw data without any analysis. The rectangular profile of the excess in this region contradicts the spherical morphology expected for the other hypotheses.
                                      
\section{Conclusion}
\label{conc}
The only signals from the existence of DM have an gravitational origin, from the flat rotation curves in present galaxies, to the fast formation of Galaxies and the power spectra of their distributions and the acoustic peaks in the CMB in the early universe. These gravitational interactions suggest that the DM consists of particles, just like normal matter, but having only weak interactions, as demonstrated by the large halos of DM around the Galaxies leading to the flat rotation curves. If the DM particles are thermal relics from the early universe, they are expected to annihilate with cross sections given in Fig. \ref{f2}. Since the annihilation cross section is 10 orders of magnitude higher than the upper limit on the scattering cross section of DM particles on nuclei, a reasonable explanation is that  the interactions between DM and visible matter are dominated by the exchange of Higgs particles, as e.g. strongly preferred in the NMSSM model of Supersymmetry. But in this case the DM particle is singlino-like, in which case the cross sections may be below the ''neutrino floor'' of the inevitably large background from cosmogenic neutrinos. For the much debated signals from DM annihilation alternative explanations exist. E.g. the positron excess can be explained by  positrons from nearby pulsars or triplet production in nearby sources, while the Fermi GeV excess of diffuse gamma-rays is best described by the emissivity of molecular clouds. For the observed annual modulation by the DAMA/LIBRA collaboration one would  like to see a confirmation by the proposed future experiments given the inconsistency with other experiments. So no convincing DM signals have been observed so far. Does this mean that our supersymmetric paradigm of how to search for DM is wrong? At present we cannot decide, since on the one hand particles different from WIMPs may constitute DM, but on the other hand the signal rates predicted by Supersymmetry are below the discovery potential of present experiments. So we have to wait for data of forthcoming experiments, like the Darwin detector with 40 tonnes of liquid Xenon for the direct DM searches, which has a sensitivity down to the neutrino floor. Or the ground-based Cherenkov Telescope Array (CTA) project, searching for annihilation signals in the gamma-ray sky with a large solid angle and excellent angular resolution \cite{Acharya:2017ttl}. The prospects for finding supersymmetric DM at the LHC are basically restricted to low masses, because the coupling of the LSP to protons is small given the stringent limits on the scattering cross section between DM and nuclei.  At the LHC the discovery potential of the Supersymmetry  is larger by searching for an extended Higgs sector. E.g. in the NMSSM one might   find an additional Higgs boson with a mass below the 125 GeV of the already discovered Higgs boson \cite{Beskidt:2017dil}. So let's wait and see.
\subsection*{Acknowledgements}
I thank my close collaborators  Conny Beskidt, Leo Bosse, Peter L. Biermann, Iris Gebauer and Dmitri Kazakov for fruitful discussions and warmly acknowledge the financial support from the Deutsche Forschungsgemeinschaft  (DFG, Grant BO 1604/3-1).  
%

\end{document}